\documentclass[letterpaper]{article} 
\usepackage{aaai2027}  
\usepackage[hyphens]{url}  
\usepackage{graphicx} 
\usepackage{natbib}  
\usepackage{caption} 
\usepackage{algorithm}
\usepackage{algorithmic}
\usepackage{booktabs}
\usepackage{multirow}
\usepackage{colortbl}
\usepackage{enumitem}
\usepackage{xcolor}
\usepackage{pdfpages}
\usepackage{amsmath}
\usepackage{graphicx}

\definecolor{background_gray}{gray}{0.9}
\usepackage{amssymb}

\usepackage{amsthm}             

\usepackage{newfloat}
\usepackage{listings}
\DeclareCaptionStyle{ruled}{labelfont=normalfont,labelsep=colon,strut=off} 
\floatstyle{ruled}
\newfloat{listing}{tb}{lst}{}
\floatname{listing}{Listing}

\usepackage{booktabs}

\title{MemoryCPT: An End-to-End Agent Memory Framework for Cost-Performance Trade-off}
\author {
    Songxin Lei\textsuperscript{\rm 1,\rm 2}\equalcontrib,
    Kun Ouyang\textsuperscript{\rm 2}\equalcontrib,
    Weilin Ruan\textsuperscript{\rm 4}, 
    Yuqian Wu\textsuperscript{\rm 3}, \\
    Zhijiang Guo\textsuperscript{\rm 1, \rm 3},
    Yushi Sun\textsuperscript{\rm 2}\corresponding,
    Fugee Tsung\textsuperscript{\rm 1}
}
\affiliations {
    \textsuperscript{\rm 1}The Hong Kong University of Science and Technology, Hong Kong SAR, China\\
    \textsuperscript{\rm 2}LIGHTSPEED STUDIOS, Tencent\\
    \textsuperscript{\rm 3}The Hong Kong University of Science and Technology (Guangzhou), Guangzhou, China\\
    \textsuperscript{\rm 4}The Chinese University of Hong Kong, Hong Kong SAR, China\\
    slei924@connect.ust.hk,
    ouyangkun@u.nus.edu,
    wruan792@connect.hkust-gz.edu.cn, \\
    ywu188@connect.hkust-gz.edu.cn, 
    zhijiangguo@hkust-gz.edu.cn, 
    ysunbp@connect.ust.hk, 
    season@ust.hk
}

\begin{document}

\maketitle

\begin{abstract}
Long-horizon LLM agents require memory systems that recover useful evidence from large interaction histories without passing excessive context to downstream models. Existing memory pipelines often rely on hand-crafted heuristics and repeated LLM calls, which can introduce redundant context and high inference cost. We propose \textbf{MemoryCPT}, an end-to-end trainable agent memory pipeline that spans offline memory construction and online query-conditioned context generation. MemoryCPT consists of two stages: \textbf{Query-agnostic Distillation (QAD)}, which distills a modular memory-construction pipeline into a compact model using explicit reasoning traces; and \textbf{Query-aware Retrieval and Summarization (QAR)}, which combines reciprocal rank fusion (RRF) with a LoRA-based summarizer trained via Group Relative Policy Optimization (GRPO) under a cost-aware reward. We further introduce \textbf{Quality per Cost (QPC)} to quantify answer quality per unit inference cost. Experiments on LoCoMo and LongMemEval show that MemoryCPT improves the cost-performance trade-off over the evaluated baselines, while ablation and sensitivity analyses characterize the contributions of its components and the effects of key design choices.
\end{abstract}


\section{Introduction}



Long-horizon LLM agents are increasingly expected to maintain persistent memory across extended interactions, including multi-session conversations, task logs, and evolving personal knowledge~\cite{zhong2023memory, shinn2023reflexion}. Unlike short-context question answering, these settings require the agent to recover evidence from a large and continuously growing history before producing a response. A straightforward solution is to feed more history into the downstream LLM, but this quickly becomes impractical: long contexts are expensive, slow, and often noisy~\cite{liu2023lost}, while useful evidence may occupy only a small fraction of the total tokens. Therefore, an effective agent memory system should not only improve answer quality, but also maximize the intelligence yielded per unit of inference cost.

Existing memory systems have made important progress by improving different parts of the context-management pipeline. Current paradigms generally fall into three categories: compressing raw interaction histories into condensed summaries~\cite{kim2026mem}, organizing memories into structured blocks or episodic records~\cite{milosevic2026episodic}, and enhancing retrieval mechanisms via dense, sparse, or hybrid ranking signals~\cite{nie2026evo}. While these training-free pipelines offer deployment flexibility, they predominantly rely on hand-crafted prompts, rigid heuristics, or external LLM calls during inference~\cite{jiang2026anatomy}. Such over-conservative context selection can substantially increase inference cost without yielding commensurate gains in response quality. This bottleneck motivates an end-to-end trainable memory framework that distills test-time memory operations into model parameters and explicitly optimizes the cost-performance trade-off. However, realizing such a paradigm presents two primary challenges.

The first challenge lies in the \textit{compression stage}: how can a compact model learn to construct high-quality long-term memories from raw histories? Unlike simple summarization~\cite{xu2026context}, a strong memory pipeline must segment raw conversations, generate episodic records, decide whether new events should be merged with existing memories, and extract persistent semantic knowledge~\cite{ji2026memory}. Directly supervising the student with only final summaries would hide these intermediate decisions and make the learning target too rough~\cite{griot2026compress}. Therefore, an end-to-end framework needs a way to decompose the compression process into learnable steps while preserving the teacher's reasoning behind each memory operation.

The second challenge lies in the \textit{extraction stage}: how can the system retrieve and compress query-relevant memories without sacrificing answer quality? Retrieving more memory records generally improves recall, but it also increases the number of tokens consumed by the final QA model~\cite{du2025context}. Conversely, overly aggressive compression reduces cost but may discard critical evidence. This creates a fundamental tension between answer quality and inference cost. Optimizing accuracy alone encourages large memory contexts, while optimizing cost alone leads to poor summaries~\cite{stiennon2022learning}. Therefore, a practical memory system must learn a query-aware policy that preserves critical evidence while minimizing redundant tokens.

To tackle these challenges, we propose \textbf{MemoryCPT}, an end-to-end trainable agent memory pipeline for cost-performance trade-off optimization. To address the compression challenge, MemoryCPT introduces \textbf{Query-agnostic Distillation (QAD)}, which distills a modular memory pipeline into a compact base model via Low-Rank Adaptation (LoRA) supervised fine-tuning (SFT). Instead of learning only final summaries, QAD trains on explicit reasoning traces and structured outputs, enabling the model to internalize reusable memory-construction skills. To address the extraction challenge, MemoryCPT further introduces \textbf{Query-aware Retrieval and Summarization (QAR)}, which first uses RRF-based retrieval for rough selection and then trains a LoRA-based summarization policy with GRPO. Its reward balances answer quality against token consumption, allowing the framework to optimize the cost-performance trade-off rather than raw accuracy alone.

Our contributions are summarized as follows:
\vspace{-0.3em}
\begin{itemize}[leftmargin=*]
    \item \textit{An end-to-end memory pipeline for cost-performance optimization.} We introduce MemoryCPT, which spans the complete memory process from query-agnostic construction to query-aware retrieval and summarization, and explicitly targets the cost-performance trade-off in long-horizon multi-turn dialogue memory.

    \item \textit{A two-stage post-training algorithm.} We develop a sequential training scheme in which QAD distills modular memory-construction skills from teacher reasoning traces, while QAR trains a query-aware summarization policy with GRPO and a cost-aware reward. This design separates memory construction from online context optimization while keeping the downstream answer model frozen.

    \item \textit{Empirical evaluation of quality and cost.} Experiments on LoCoMo and LongMemEval compare MemoryCPT with representative baselines under a unified evaluation pipeline. Main results, component ablations, and sensitivity analyses examine the quality-cost trade-off and the effects of the reward coefficient and retrieval depth.
\end{itemize}

\section{Problem Statement}

We formulate long-horizon agent memory from a cost-performance perspective: a memory system should improve answer quality while minimizing inference cost.

\noindent\textbf{Long-horizon memory QA.}
Let $\mathcal{H}=\{h_i\}_{i=1}^{N}$ denote an agent's historical records and $q$ a user query. Since feeding the full history into an LLM is costly and often noisy, a memory system first derives a compact query-relevant memory $m_q$ and then answers with a fixed QA model:
\begin{equation}
    m_q = f_{\mathrm{mem}}(q, \mathcal{H}), \qquad
    \hat{y} = f_{\mathrm{ans}}(q, m_q),
    \label{eq:memory_qa}
\end{equation}
where $\hat{y}$ is the predicted answer and $y^*$ is the reference answer. MemoryCPT instantiates $f_{\mathrm{mem}}$ with two trainable stages:
\begin{equation}
    \mathcal{M} = f_{\mathrm{QAD}}(\mathcal{H}), \qquad
    m_q = f_{\mathrm{QAR}}(q, \mathcal{M}),
    \label{eq:two_stage_memory}
\end{equation}
where Query-agnostic Distillation (QAD) constructs reusable episodic and semantic memories $\mathcal{M}$ offline, and Query-aware Retrieval and Summarization (QAR) selects and compresses query-relevant memories online.

\noindent\textbf{Cost and Quality per Cost.}
To ensure fair comparison, the cost of answering $q$ must account for both online operations and the amortized offline processing. Let $n$ be the total number of queries over the history $\mathcal{H}$. The amortized cost per query is defined as:
\begin{equation}
    C_{\mathrm{amortized}}(n) = \frac{C_{\mathrm{QAD}}}{n} + C_{\mathrm{QAR+QA}},
    \label{eq:cost_amortized}
\end{equation}
where $C_{\mathrm{QAD}}$ represents the offline memory construction cost, and $C_{\mathrm{QAR+QA}}$ is the online cost covering memory processing and final QA. We compute the monetary cost for any phase by separately counting input/output tokens for each LLM call:
\begin{equation}
    \mathcal{C} = \sum_{j \in \mathcal{J}}
    \left( p^{\mathrm{in}}_j T^{\mathrm{in}}_j
    + p^{\mathrm{out}}_j T^{\mathrm{out}}_j \right),
    \label{eq:cost}
\end{equation}
where $\mathcal{J}$ is the set of model calls for that phase, $T_j^{\mathrm{in/out}}$ are token counts, and $p_j^{\mathrm{in/out}}$ are token prices. To evaluate memory systems beyond raw quality, we define \textbf{Quality per Cost (QPC)} as quality per unit cost:
\begin{equation}
    \mathrm{QPC} = \frac{Q(\hat{y}, y^*)}{C_{\mathrm{amortized}}(n)}.
    \label{eq:QPC_general}
\end{equation}
In experiments, $Q$ is token-level F1 and Cost is reported as USD per query multiplied by $10^4$:
\begin{equation}
    \mathrm{QPC} = \frac{\mathrm{F1}(\hat{y}, y^*)}{\mathrm{Cost}(q)}.
    \label{eq:QPC_f1}
\end{equation}
A higher QPC indicates that the system obtains more correct answer information under the same amortized budget.

\noindent\textbf{Optimization objective.}
Given queries $\mathcal{Q}$, our goal is to learn memory parameters $\theta$ that maximize expected QPC:
\begin{equation}
    \theta^* = \arg\max_{\theta}
    \mathbb{E}_{q \sim \mathcal{Q}}
    \left[
    \frac{Q\bigl(f_{\mathrm{ans}}(q, f_{\mathrm{mem},\theta}(q, \mathcal{H})), y^*\bigr)}
    {C_{\mathrm{amortized},\theta}(n)}
    \right].
    \label{eq:objective}
\end{equation}
This objective captures the core challenge of MemoryCPT: retaining enough evidence for correct answers while avoiding redundant tokens in both offline construction and expensive online downstream LLM calls.

\section{Methodology}
\label{sec:methodology}

\subsection{Framework Overview}
\label{sec:framework_overview}

\begin{figure*}[t]
    \centering
    \includegraphics[width=0.95\textwidth]{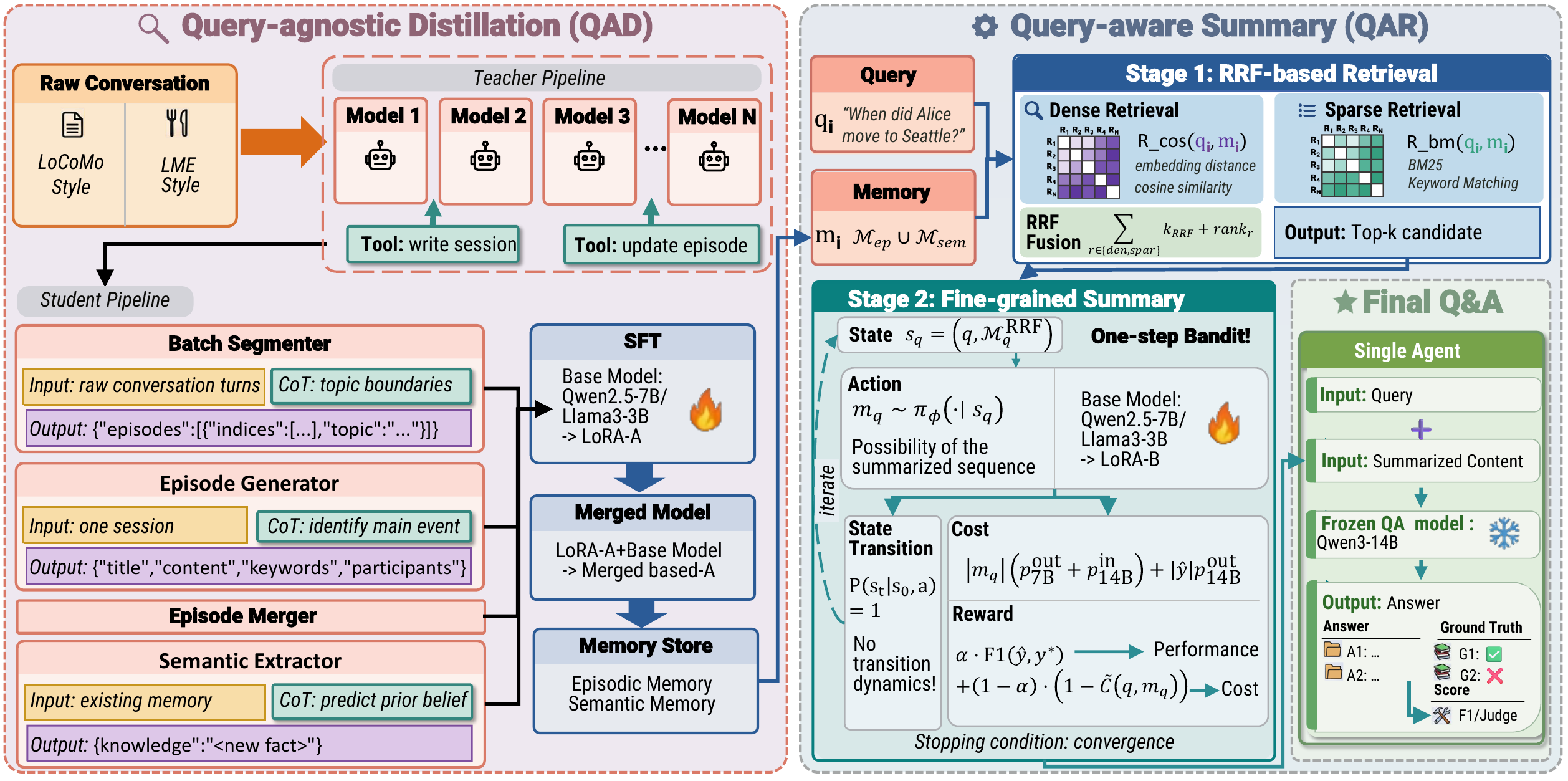}
    \caption{Overview of MemoryCPT. The offline stage, Query-agnostic Distillation (QAD), distills a modular memory pipeline into LoRA-A and constructs a reusable memory store with episodic and semantic memories. The online stage, Query-aware Retrieval and Summarization (QAR), first applies RRF-based retrieval as rough selection and then uses a GRPO-trained LoRA-B summarizer to generate a query-conditional fine summary $m_q$. The final QA model is frozen. Dashed arrows indicate training-time reward feedback.}
    \label{fig:framework}
    \vspace{-1.5em}
\end{figure*}

MemoryCPT is an end-to-end trainable memory pipeline that maps raw interaction histories to query-conditioned contexts via two sequential stages (Figure~\ref{fig:framework}):
\begin{equation}
    \mathcal{H} \xrightarrow{f_{\text{QAD}}} \mathcal{M}, \quad
    (q, \mathcal{M}) \xrightarrow{f_{\text{QAR}}} m_q, \quad
    (q, m_q) \xrightarrow{f_{\text{ans}}} \hat{y}.
    \label{eq:method_overview}
\end{equation}
Only Query-agnostic Distillation ($f_{\text{QAD}}$) and Query-aware Retrieval and Summarization ($f_{\text{QAR}}$) are trainable; the downstream QA model ($f_{\text{ans}}$) remains frozen. 

Specifically, \textbf{QAD} runs offline to distill raw dialogues $\mathcal{H}$ into reusable episodic and semantic memories $\mathcal{M}$ using teacher traces~\cite{ma2026nemori}. Online, \textbf{QAR} processes a user query $q$ by retrieving candidate memories via Reciprocal Rank Fusion (RRF) and compressing them into a compact summary $m_q$ via a GRPO-trained summarizer. This offline-online separation isolates memory construction from query-specific extraction, reducing the RL action space and amortizing expensive construction costs across queries.

\subsection{Query-agnostic Distillation}
\label{sec:QAD}

QAD distills a teacher's memory-construction capabilities into a compact student via a modular pipeline~\cite{ma2026nemori}, learning from role-level reasoning traces rather than direct end-to-end mapping.

\subsubsection{Modular Distillation Pipeline.}
The teacher decomposes memory construction into four roles: \textbf{BatchSegmenter} (groups conversation turns), \textbf{EpisodeGenerator} (creates episodic records), \textbf{EpisodeMerger} (merges related episodes), and \textbf{SemanticExtractor} (extracts persistent knowledge). This modularity provides hardness scaffolding, breaking a complex mapping task into self-contained subtasks with well-defined intermediate targets.

We collect role-specific teacher outputs in ShareGPT format: each sample pairs a prompt $x^\rho$ with a reasoning trace $r^\rho$ and a structured JSON output $y^\rho$. The student, Qwen2.5-7B-Instruct~\cite{yang2024qwen2_5} with a LoRA adapter (LoRA-A), is supervised on these targets. Post-training, LoRA-A is merged into the base model to yield a standalone checkpoint $base_A$.

\subsubsection{Reasoning Trace Construction.}
Teacher responses follow a rationale-then-output format, providing explicit reasoning within \texttt{<think>...</think>} tags before outputting executable JSON. This dense supervision teaches the student not just \textit{what} memory operations to execute, but \textit{why}. To reduce inference overhead, adjacent teacher calls (e.g., merge decision and content generation) are collapsed into single student predictions. The target concatenates the reasoning and output ($z^\rho = r^\rho \oplus y^\rho$), optimized via standard causal language modeling:
\begin{equation}
    \mathcal{L}_{\text{QAD}}
    = - \mathbb{E}_{(x,r,y) \sim \mathcal{D}_{\text{QAD}}}
    \sum_{t=1}^{|z|} \log p_\theta(z_t \mid z_{<t}, x),
    \quad z = r \oplus y,
    \label{eq:QAD_loss}
\end{equation}
where $\theta$ denotes the LoRA-A parameters over a frozen base model.

\subsubsection{Memory Operations as Structured Actions.}
The student's structured JSON outputs serve as discrete operations executed by an external runtime, which handles database management. This abstracts the final memory store into $\mathcal{M}=\mathcal{M}_{\text{ep}} \cup \mathcal{M}_{\text{sem}}$ (episodic and semantic memories), ensuring the model learns explicit decision-making—when to create, merge, or abstract events—rather than generating a monolithic unstructured text summary.

\subsection{Query-aware Retrieval and Summarization}
\label{sec:qar}

QAR is the online component of MemoryCPT. Given a query $q$ and memory store $\mathcal{M}$, it performs rough retrieval to bound the candidate set, then compresses these candidates into a query-aware summary $m_q$, reducing search space and optimizing the content-cost trade-off.

\subsubsection{RRF-based Retrieval.}
Retrieval uses dense (embedding) and sparse (BM25) signals over episodic and semantic memories. To reconcile their different score scales, we fuse ranks via Reciprocal Rank Fusion (RRF)~\cite{Rackauckas2024med}:
\begin{equation}
    s_{\text{RRF}}(m;q)
    = \sum_{r \in \{\text{dense},\text{sparse}\}}
    \frac{1}{k_{\text{rrf}} + \operatorname{rank}_r(m \mid q)},
    \label{eq:rrf}
\end{equation}
where $\operatorname{rank}_r$ is the retriever rank and $k_{\text{rrf}}$ is a smoothing constant. RRF is applied separately to episodic and semantic indices, concatenating top candidates to form $\mathcal{M}^{\text{RRF}}_q$. This rough selection bounds the input, sparing the summarizer from attending to the entire store and significantly reducing the effective action space.

\subsubsection{Policy, State, and Action.}
We formulate QAR training as a contextual bandit problem. The state pairs the query with RRF-selected candidates:
\begin{equation}
    s_q = (q, \mathcal{M}^{\text{RRF}}_q).
    \label{eq:bandit_state}
\end{equation}
The action is the summary token sequence $m_q$ generated in one autoregressive rollout:
\begin{equation}
    m_q \sim \pi_\phi(\cdot \mid s_q),
    \label{eq:policy_action}
\end{equation}
where $\pi_\phi$ uses $base_A$ equipped with a trainable LoRA-B adapter ($\phi$). The frozen QA model then consumes $(q,m_q)$ to output $\hat{y}$. We optimize LoRA-B via Group Relative Policy Optimization (GRPO)~\cite{shao2024med}:
\begin{equation}
    \max_{\phi}\; \mathbb{E}_{q}\,\mathbb{E}_{m_q \sim \pi_\phi(\cdot \mid s_q)}
    \left[ R(q,m_q) \right]
    - \beta D_{\text{KL}}\!\left(\pi_\phi \Vert \pi_{\text{ref}}\right),
    \label{eq:grpo_objective}
\end{equation}
where $R(q,m_q)$ is the reward. GRPO normalizes rollout rewards against a sampled group mean to compute relative advantage.

\subsubsection{Cost Modeling.}
The reward utilizes a training-time proxy for variable online marginal costs. Given summary $m_q$ and answer $\hat{y}$, the variable online cost is:
\begin{equation}
    C(q,m_q)
    = |m_q|\left(p_{\text{7B}}^{\text{out}} + p_{\text{14B}}^{\text{in}}\right)
    + |\hat{y}|p_{\text{14B}}^{\text{out}},
    \label{eq:training_cost}
\end{equation}
where $|m_q|$ and $|\hat{y}|$ are token counts, and $p$ denotes respective token prices. Fixed costs (e.g., database retrieval) are excluded from this training proxy as they cancel during GRPO advantage estimation. We normalize the training proxy cost to $[0,1]$ via clipping:
\begin{equation}
    \widetilde{C}(q,m_q) = \min\left(1, \frac{C(q,m_q)}{C_{\text{ref}}}\right).
    \label{eq:cost_norm}
\end{equation}
This bounds the proxy and aligns its scale with F1.

\subsubsection{Reward Design.}
The reward combines F1 quality (computed via the frozen QA model against gold answer $y^*$) and cost saving:
\begin{equation}
    R(q,m_q)
    = \alpha \cdot \mathrm{F1}(\hat{y}, y^*)
    + (1-\alpha) \cdot \left(1 - \widetilde{C}(q,m_q)\right),
    \label{eq:reward}
\end{equation}
where $\alpha \in [0,1]$ balances quality and cost. We use token-level F1 rather than an LLM judge for training because it is continuous and locally computable, offering dense feedback. Instead of directly optimizing the potentially unstable ratio $\mathrm{QPC} = \mathrm{F1}/\mathrm{Cost}$, we use this weighted-sum surrogate. Both terms are bounded in $[0,1]$, ensuring stable reward magnitudes.

\subsection{End-to-end Memory Pipeline}
\label{sec:end_to_end_framework}

\subsubsection{Two-stage Training Phase.}
MemoryCPT is trained with two LoRA adapters that serve different purposes. \textbf{LoRA-A} is trained first with QAD. The base Qwen2.5-7B-Instruct model is frozen, and only LoRA-A is updated using the SFT objective in Equation~\ref{eq:QAD_loss}. After training, LoRA-A is merged into the base model to produce $base_A$, a standalone model that can build the memory store offline.

\textbf{LoRA-B} is then trained on top of $base_A$ for QAR. For each training query, we run RRF-based retrieval over the memory store to obtain $\mathcal{M}^{\text{RRF}}_q$, use $(q,\mathcal{M}^{\text{RRF}}_q)$ as the policy input, and optimize LoRA-B with GRPO using the reward in Equation~\ref{eq:reward}. LoRA-B is newly initialized rather than continued from LoRA-A; $base_A$ remains frozen during GRPO. Thus, LoRA-A encodes query-agnostic memory-construction skills, while LoRA-B learns query-aware cost-performance optimization.

The full training flow is as follows. First, the teacher pipeline produces role-level reasoning traces and structured JSON outputs. Second, these outputs are converted into single-turn SFT data and used to train LoRA-A. Third, LoRA-A is merged into the base model to obtain $base_A$, which constructs the episodic and semantic memory store. Finally, RRF retrieves candidate memories for each training query, and LoRA-B is trained with GRPO to generate cost-aware query summaries.

\subsubsection{Inference Phase.}
At inference time, QAD has already been completed offline. The online pipeline therefore contains only three steps:
\begin{equation}
\begin{aligned}
    \mathcal{M}^{\text{RRF}}_q &= \mathrm{RRF}(q,\mathcal{M}), \\
    m_q &\sim \pi_\phi(\cdot \mid q,\mathcal{M}^{\text{RRF}}_q), \\
    \hat{y} &= f_{\text{ans}}(q,m_q).
\end{aligned}
    \label{eq:inference_pipeline}
\end{equation}
First, RRF retrieves a bounded set of episodic and semantic candidates from the pre-built memory store. Second, the LoRA-B summarizer generates a single query-aware summary. Third, the frozen QA model reads the query and summary to produce the final answer. Unlike training, inference does not sample multiple rollouts, does not use gold answers, and does not invoke the reward model.

For evaluation, we compare $\hat{y}$ with the gold answer $y^*$ to compute token-level F1, invoke an LLM judge for semantic correctness, and log token counts across all inference-time model calls to compute Cost and QPC. These evaluation components are not part of the deployed inference loop.

\section{Experiments}

We evaluate MemoryCPT from a cost-performance perspective. Our main question is whether an end-to-end trainable memory pipeline spanning offline construction and online query-conditioned summarization can improve answer quality while reducing inference-time cost. 


\subsection{Experimental Settings}
\subsubsection{Datasets \& Evaluation Metrics.}

We evaluate on two long-horizon dialogue memory benchmarks, LoCoMo~\cite{maharana2024ben} and LongMemEval~\cite{wu2025ben}. 
Following our memory-focused setting, we evaluate on Cat1--Cat4 and exclude Cat5. We use conv-49 and conv-50 as the test split, resulting in 314 test questions, and use the remaining eight conversations for training the online memory policy. 
Since LongMemEval is released as a test-only benchmark, we construct a stratified split with 150/98/105 questions for train/validation/test, preserving the distribution of its six memory categories.
\begin{table}[t]
\centering
\setlength{\tabcolsep}{3.5pt}
\renewcommand{\arraystretch}{1.12}
\resizebox{\columnwidth}{!}{%
\begin{tabular}{@{}lcccc@{}}
\toprule
\textbf{Dataset} & \textbf{Train} & \textbf{Val} & \textbf{Test} & \textbf{Avg. Context} \\
\midrule
LoCoMo & 8 convs & -- & 314 QAs & 8.27K tokens \\
LongMemEval-S & 150 QAs & 98 QAs & 105 QAs & 9.37K tokens \\
\bottomrule
\end{tabular}%
}
\caption{Dataset statistics. LoCoMo is evaluated on Cat1--Cat4 questions from conv-49/50.} 
\label{tab:datasets}
\end{table}
We then report four metrics: (i) \textbf{F1}~\citep{maharana2024ben};(ii) \textbf{LLM-as-Judge(Judge)}; 
(iii)\textbf{Cost} measures the average inference-time monetary cost per question. It covers all inference-time LLM calls in memory construction, retrieval, summary, and final QA, distinguishes input and output tokens, and uses a unified price table for all methods. For readability, we report Cost as USD per question multiplied by $10^4$. Finally, we define \textbf{Quality per Cost (QPC)} as
\begin{equation}
    \mathrm{QPC} = \frac{\mathrm{F1}}{\mathrm{Cost}},
\end{equation}
where Cost is the reported value in USD per question multiplied by $10^4$. QPC therefore measures the answer quality obtained per unit inference cost; higher values indicate a better cost-performance trade-off.



\subsubsection{Baselines.}
 
\begin{table*}[!htb]
\centering
\setlength{\tabcolsep}{4.5pt}
\renewcommand{\arraystretch}{1.18}
\resizebox{\textwidth}{!}{%
\begin{tabular}{@{}l cccc cccc cccc@{}}
\toprule
 & \multicolumn{4}{c}{\textbf{LoCoMo}} & \multicolumn{4}{c}{\textbf{LongMemEval}} & \multicolumn{4}{c}{\textbf{Avg.}} \\
\cmidrule(lr){2-5} \cmidrule(lr){6-9} \cmidrule(lr){10-13} 
\textbf{Model} 
  & F1$\uparrow$ & Judge$\uparrow$ & Cost$\downarrow$ & QPC$\uparrow$
  & F1$\uparrow$ & Judge$\uparrow$ & Cost$\downarrow$ & QPC$\uparrow$
  & F1$\uparrow$ & Judge$\uparrow$ & Cost$\downarrow$ & QPC$\uparrow$ \\
\midrule
\rowcolor{gray!25}
\multicolumn{13}{@{}l}{\textit{Reference open-source algorithms (not bolded)}} \\
Qwen-3-14B (no memory)      & 0.319 & 0.557 & 11.11 & 0.029 & 0.118  & 0.162  & 11.11  & 0.011 & 0.219 & 0.360 & 11.11 & 0.020\\
Budgetmem~\cite{zhang2026mem}   & 0.373 & 0.640 & 24.11 & 0.015 & 0.358 & 0.467 & 18.1 & 0.020 & 0.366 & 0.554 & 21.10 & 0.018\\
\midrule
\rowcolor{gray!25}
Qwen-2.5-7B (Base)~\cite{qwen2.5}      &  &  &  &  &   &   &   &  &  &  &  & \\
\quad + LightMem~\cite{fang2026mem}  & 0.433 & 0.755 & 19.80 & 0.022  & 0.334 & 0.457 & 63.02 & 0.005 & 0.384 & 0.606 & 41.41 & 0.009 \\
\quad + MemoryOS~\cite{kang2025mem}   & 0.354 & 0.573 & 25.23 &  0.014 & 0.386 & 0.486 & 396.31 & 0.001 & 0.370 & 0.530 & 210.77 & 0.002 \\
\quad + Memory-R1~\cite{yan2025memory}   & 0.372 & 0.592 & \textbf{3.78} &  0.098 & 0.231 & 0.267 & 5.96 & 0.039 & 0.302 & 0.430 & \textbf{4.87} & 0.062 \\
\rowcolor{gray!10}
\quad + MemoryCPT (Ours) $\bigstar$  & \textbf{0.479} & \textbf{0.755} & 4.31 & \textbf{0.111} & \textbf{0.485} & \textbf{0.533} & \textbf{5.72} & \textbf{0.085} & \textbf{0.482} & \textbf{0.644} & 5.02 & \textbf{0.096}\\
\midrule
\rowcolor{gray!25}
Llama-3.2-3B (Base)~\cite{grattafiori2024llama3}      &  &  &  &  &   &   &   &  &  &  &  & \\
\quad + LightMem~\cite{fang2026mem}  &  0.372 & 0.640 & 11.73 & 0.032 & \textbf{0.293} & \textbf{0.391} & 43.06 & 0.007 & \textbf{0.333} & \textbf{0.516} & 27.40 & 0.012 \\
\quad + MemoryOS~\cite{kang2025mem}   & 0.312 & 0.538 & 10.63 & 0.029 & 0.276 & 0.290 & 159.07  & 0.002 & 0.294 & 0.414 & 84.85 & 0.003\\
\quad + Memory-R1~\cite{yan2025memory}   & 0.326 & 0.576 & \textbf{3.70} &  0.082 & 0.131 & 0.165 & 5.23 & 0.025 & 0.229 & 0.371 & \textbf{4.47} & 0.051 \\
\rowcolor{gray!10}
\quad + MemoryCPT (Ours) $\bigstar$  & \textbf{0.474} & \textbf{0.742} & 4.23 & \textbf{0.112} & 0.181 & 0.219 & \textbf{4.76} & \textbf{0.038} & 0.328 & 0.481 & 4.50 & \textbf{0.073}\\
\midrule

\bottomrule
\end{tabular}%
}
\caption{Main results on LoCoMo and LongMemEval. Methods are grouped by the base model used for memory processing (Qwen-2.5-7B or Llama-3.2-3B). 
F1 and Judge measure answer quality; Cost denotes the average inference cost per query, reported as USD $\times 10^4$; QPC = F1 / Cost measures quality per unit cost. Best results are in bold.}
\label{tab:main_accuracy}
\end{table*}

We compare MemoryCPT with five baselines. \textbf{No-Memory} uses Qwen3-14B to directly read the raw multi-session conversation truncated to the context limit, without memory extraction, retrieval, or summarization. \textbf{LightMem}~\cite{fang2026mem} is a lightweight training-free memory pipeline that stores and retrieves compact turn-level memories. \textbf{MemoryOS}~\cite{kang2025mem} organizes long-term conversation histories into hierarchical memory blocks with structured read/write operations. \textbf{BudgetMem}~\cite{zhang2026mem} is a budget-aware training-based memory pipeline that dynamically allocates LLM computation for memory processing and QA. \textbf{Memory-R1}~\cite{yan2025memory} is a reinforcement learning framework that equips LLMs with specialized agents for adaptive memory management and reasoning.

For the baselines we implement, we use the same final QA model, Qwen3-14B, with temperature 0 and the same prompt template. Specifically for BudgetMem, due to its inherent architectural design that prevents substituting its base model, we directly utilize the Qwen-2.5-72B API to execute its intermediate memory management operations. We also use the same Qwen2.5-72B-Instruct judge, the same test questions, and the same tokenizer and price table for cost accounting. This protocol allows us to compare memory systems by the same downstream objective: answer quality per unit inference cost.

\subsubsection{Implementation Details.}

All experiments run on a single node with 4 NVIDIA H20 GPUs. We use LLaMA-Factory for LoRA SFT and vLLM for high-throughput inference. MemoryCPT uses Qwen2.5-7B-Instruct and Llama-3.2-3B for memory operations and Qwen3-14B for final QA. The LLM judge is Qwen2.5-72B-Instruct.

MemoryCPT is trained in two stages. In Stage 1, \textbf{Query-agnostic Distillation (QAD)} distills memory extraction traces into a LoRA adapter on top of Qwen2.5-7B-Instruct. The training data contains 1185/257 train/validation examples for LoCoMo and 4766/239 train/validation examples for LongMemEval. We train LoRA-A with rank 16, alpha 32, dropout 0.05, learning rate $1\times10^{-4}$, cutoff length 8192, and select the checkpoint with the best validation loss. After training, LoRA-A is merged into the base model to obtain \texttt{base\_A}. In Stage 2, \textbf{Query-aware Retrieval and Summarization (QAR)} first performs an RRF-based rough selection, and then applies \textbf{Fine Summary (FS)}, a GRPO-trained LoRA-B on top of \texttt{base\_A}, to produce query-conditional summaries. We set the main reward coefficient to $\alpha=0.8$ and defer the full alpha sweep to a later experiment.

At inference time, a query first retrieves memory chunks via reciprocal rank fusion over dense Qwen3-Embedding retrieval and sparse BM25 retrieval. We retrieve top-20 episodic memories and top-50 semantic memories before fusion. The retrieved memories are then compressed by \texttt{base\_A}+LoRA-B into a short query-aware summary, which is finally read by Qwen3-14B for answer generation. 

\subsection{Model Comparison}




\textbf{Table~\ref{tab:main_accuracy}} shows that MemoryCPT achieves the strongest cost-performance trade-off. Using Qwen-2.5-7B as the base model, compared with BudgetMem,, MemoryCPT improves LoCoMo F1 from 0.373 to 0.479 and Judge from 0.640 to 0.755, while reducing Cost from 24.11 to 3.46. This corresponds to around 28\% relative F1 improvement, 7 times lower inference cost, and over 9 times higher QPC. These results show that end-to-end memory training can jointly improve answer quality and inference efficiency, rather than trading one for the other.

The comparison also reveals the cost-quality bottlenecks of existing memory pipelines. On LoCoMo, LightMem, MemoryOS, and BudgetMem improve F1 over the No-Memory baseline from 0.319 to the range of 0.312 to 0.433, but their inference cost drastically increases from 11.11 to 25.23 due to expensive intermediate context processing. Conversely, while the RL-based Memory-R1 successfully restricts cost, it yields lower answer quality. MemoryCPT consistently avoids this compromise across both base model architectures, delivering superior F1 scores while maintaining minimal inference costs.

QPC makes this efficiency gap explicit. On LoCoMo with the Qwen-2.5-7B base model, MemoryCPT reaches an QPC of 0.138, whereas No-Memory, LightMem, MemoryOS, and BudgetMem achieve only 0.029, 0.022, 0.014, and 0.015, respectively. While the efficiency-optimized Memory-R1 reaches an QPC of 0.105, it still trails behind our method. This robust advantage persists when switching to the Llama-3.2-3B base model, supporting our central claim that long-horizon memory systems should be evaluated by the intelligence yielded per unit cost, not just raw accuracy.

\subsection{Ablation Study}

\textbf{Table~\ref{tab:ablation}} verifies the contribution of each component on two datasets.

\begin{table*}[!htb]
\centering
{\footnotesize

\setlength{\tabcolsep}{4.5pt}
\renewcommand{\arraystretch}{1.18}
\resizebox{\textwidth}{!}{%
\begin{tabular}{@{}l cccc cccc cccc@{}}
\toprule
\multirow{2}{*}{\textbf{Model}} & \multicolumn{4}{c}{\textbf{LoCoMo}} & \multicolumn{4}{c}{\textbf{LongMemEval}} & \multicolumn{4}{c}{\textbf{Avg.}} \\
\cmidrule(lr){2-5} \cmidrule(lr){6-9} \cmidrule(lr){10-13}
 & F1$\uparrow$ & Judge$\uparrow$ & Cost$\downarrow$ & QPC$\uparrow$
 & F1$\uparrow$ & Judge$\uparrow$ & Cost$\downarrow$ & QPC$\uparrow$
 & F1$\uparrow$ & Judge$\uparrow$ & Cost$\downarrow$ & QPC$\uparrow$ \\
\midrule
\rowcolor{gray!25}
Qwen3-14B (Base) & 0.319 & 0.557 & 11.11 & 0.029 & 0.118 & 0.162 & 11.11 & 0.011 & 0.219 & 0.360 & 11.11 & 0.020\\
w/o QAD & 0.370 & 0.583 & \textbf{2.21} & 0.131 & 0.349 & 0.457 & 9.22 & 0.038 & 0.360 & 0.520 & 5.72 & 0.063\\
w/o QAR & 0.426 & 0.745 & 11.10 & 0.038 & 0.192 & 0.248 & 11.10 & 0.017 & 0.309 & 0.497 & 11.10 & 0.028\\
w/o FS & 0.446 & \textbf{0.759} & 8.35 & 0.053 & 0.403 & 0.524 & 11.71 & 0.034 & 0.425 & 0.642 & 10.03 & 0.042 \\
\rowcolor{gray!10}
MemoryCPT (Ours) $\bigstar$ & \textbf{0.479} & 0.755 & 4.31 & \textbf{0.111} & \textbf{0.485} & \textbf{0.533} & \textbf{5.72} & \textbf{0.085} & \textbf{0.482} & \textbf{0.644} & \textbf{5.02} & \textbf{0.096}\\
\bottomrule
\end{tabular}%
}
}
\caption{Ablation study on LoCoMo and LongMemEval. QAD denotes Query-agnostic Distillation. QAR denotes Query-aware Retrieval and Summarization, consisting of RRF-based rough selection and FS. FS denotes Fine Summary, implemented by the GRPO-trained LoRA-B after RRF retrieval. Best results are in bold.}
\label{tab:ablation}
\end{table*}

\vspace{-0.3em}

\subsubsection{Effects of Query-agnostic Distillation (QAD).}

 Removing QAD means replacing our offline-distilled memory extractor with a vanilla 7B model. This reduces Cost from 3.46 to 2.21, but also drops F1 from 0.479 to 0.370 and Judge from 0.755 to 0.583. Consequently, QPC decreases from 0.138 to 0.131. This indicates that QAD acts as a quality amplifier: it introduces a small amount of additional memory-processing cost, but this cost is outweighed by substantially better downstream memory quality.

\subsubsection{Effects of Query-aware Retrieval and Summarization (w/o QAR).}

Removing QAR disables the query-aware stage after memory construction. In this case, the QA model reads a much larger retrieved memory context without query-aware compression. The result is a large increase in Cost, from 3.46 to 11.10, while F1 and Judge also fall to 0.426 and 0.745. QPC therefore collapses from 0.138 to 0.038. This confirms that QAR is essential for cost control: the rough top-$k$ selection and subsequent query-aware compression prevent memories from being passed to the QA model.

\subsubsection{Effects of Fine Summary (w/o FS).}

Finally, removing FS keeps the rough RRF-based retrieval stage but removes the GRPO-trained fine summarizer. This variant maintains a Judge score close to the full model, and even slightly higher in this run, but its Cost rises from 3.46 to 8.35 and its QPC drops from 0.138 to 0.053. We therefore do not claim that MemoryCPT is best on every individual quality metric. Instead, the key observation is that FS preserves semantic correctness while greatly improving cost efficiency. Across all ablations, only the full MemoryCPT obtains both high answer quality and the best quality-per-cost trade-off.

\subsection{Cost-Performance Trade-off Analytics}
\label{sec:cost_perf_analytics}


We finally analyze how the reward coefficient $\alpha$ in Equation~\ref{eq:reward} governs the cost-performance trade-off, and justify our default choice $\alpha=0.8$. Recall that QPC is defined as $\mathrm{F1}/\mathrm{Cost}$, where Cost aggregates all inference-time model calls. To further isolate the effect of summary length on the dominant downstream expense, we additionally report \textbf{QPC (QA)}, which keeps the same F1 but restricts Cost to the final QA call only, i.e., the token cost of feeding $(q, m_q)$ into the frozen QA model and generating $\hat{y}$. A higher QPC (QA) indicates that the summarizer delivers more answer quality per token actually consumed downstream.

\begin{figure}[!t]  
  \centering
  \includegraphics[width=0.95\linewidth]{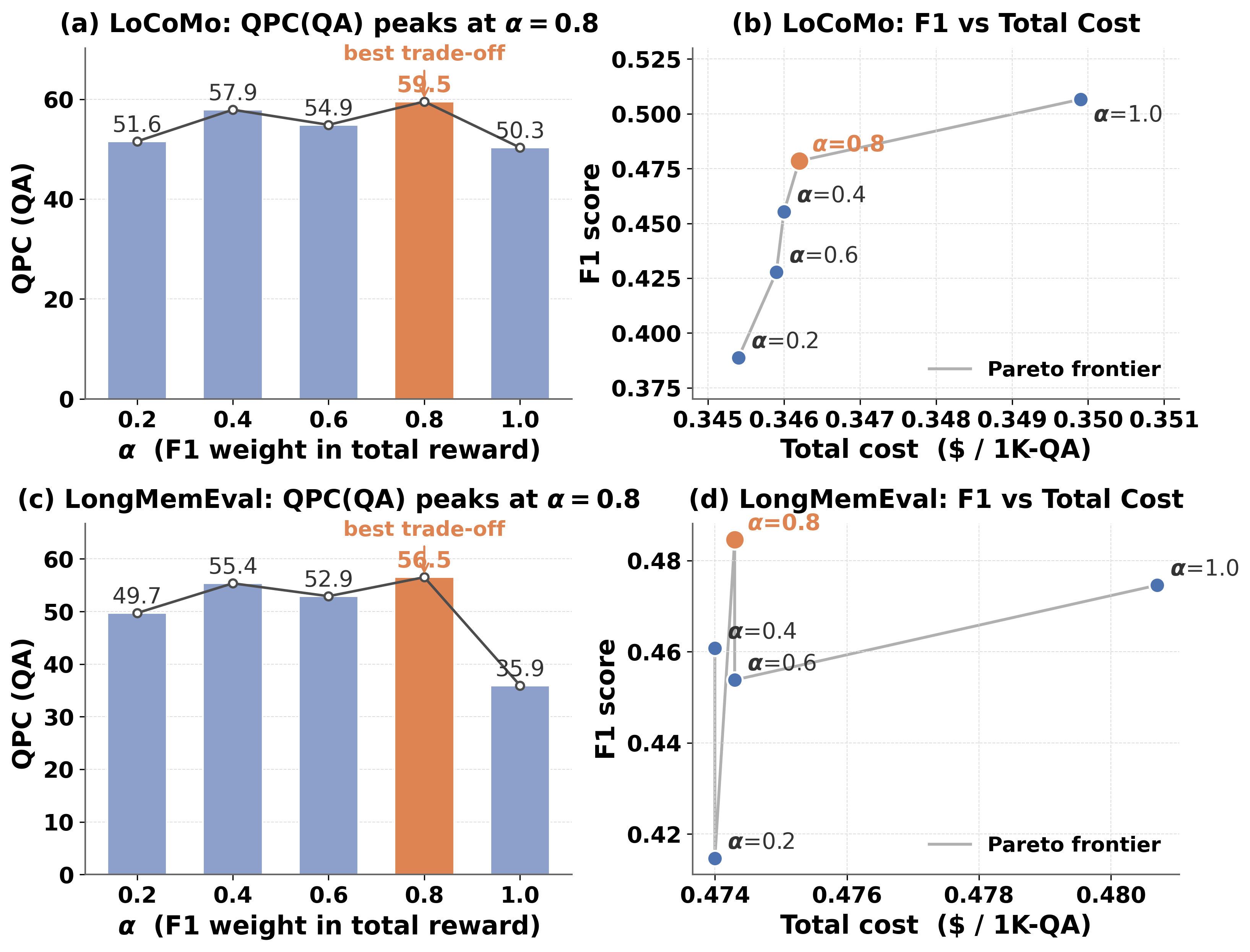}  
  \vspace{-0.5em}
  \caption{Cost-Performance Trade-off Analytics.}  
  \label{fig:convergence}
  \vspace{-0.5em}
\end{figure}

We vary $\alpha \in \{0.2, 0.4, 0.6, 0.8, 1.0\}$ and retrain LoRA-B on both datasets. \textbf{Fig.~\ref{fig:convergence}} reports two views: the bar charts show QPC (QA) as a function of $\alpha$, and the Pareto plots show the resulting F1 against total cost per query.

As shown in \textbf{Fig.~\ref{fig:convergence}}, QPC (QA) is maximized at $\alpha=0.8$ on both LoCoMo (59.5) and LongMemEval (56.5), forming a clear peak over neighboring settings. We attribute this to two opposing effects: a) When $\alpha \to 1.0$, the reward almost ignores cost, so the policy tends to produce longer summaries; F1 improves only marginally while cost rises, dragging QPC (QA) down. b) When $\alpha$ is small , the cost term dominates and the policy over-compresses, discarding answer-critical evidence and sharply reducing F1. The Pareto plots confirm this pattern: $\alpha=1.0$ sits at the high-cost end with little F1 gain, whereas aggressive compression collapses F1 at comparable cost. In contrast, $\alpha=0.8$ lies on the upper-left of the frontier, achieving near-best F1 at low cost, which is why we adopt it as the default trade-off in all main experiments.

\subsection{Retrieval Depth Analytics}
\label{sec:retrieval_depth}

We study how the retrieval depth $k$ in RRF-based rough selection affects downstream quality, and justify our default top-$20$ episodic / top-$50$ semantic candidates. A small $k$ risks dropping answer-critical evidence before summarization, whereas a large $k$ enlarges the candidate set, inflating input tokens and diluting the summarizer with irrelevant memories. We therefore sweep $k \in \{2/5, 4/10, 8/20, 20/50, 32/80\}$ (episodic/semantic) and report both F1 and LLM-Judge quality on the two datasets.

\begin{figure}[!t]
  \centering
  \includegraphics[width=0.95\linewidth]{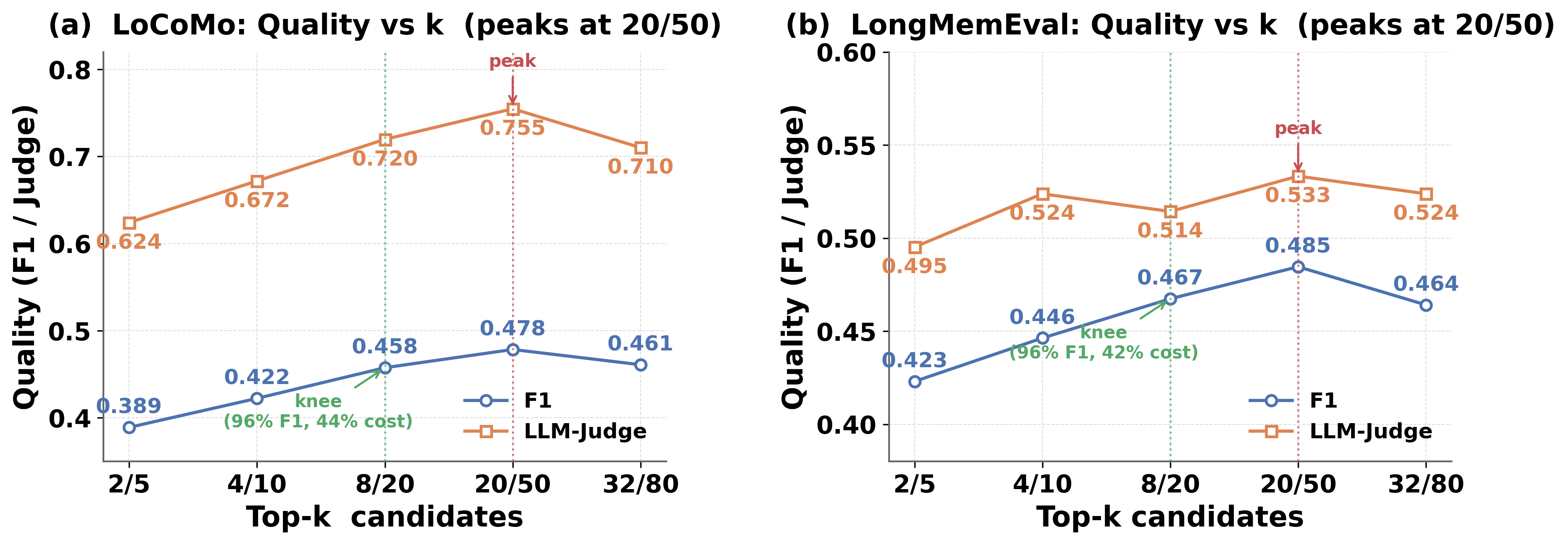}
  \vspace{-0.5em}
  \caption{Retrieval Depth Analytics.}
  \label{fig:topk}
  \vspace{-0.5em}
\end{figure}

As shown in \textbf{Fig.~\ref{fig:topk}}, both datasets exhibit the same pattern. Quality rises quickly at small $k$ and shows a knee around $8/20$, which already recovers about $96\%$ of the peak F1 at only $42$--$44\%$ of the candidate cost. Quality then peaks at $20/50$ (F1 $0.478$/Judge $0.755$ on LoCoMo, F1 $0.485$/Judge $0.533$ on LongMemEval). Beyond this point, enlarging $k$ to $32/80$ slightly degrades both metrics on both datasets, since the additional low-ranked memories introduce noise rather than new evidence. We therefore adopt $20/50$ as the default depth: it attains the best quality just before the declining regime, while keeping the candidate set compact enough for cost-efficient summarization.

\section{Related Work}

Existing work typically improves upstream context management from three perspectives: (i) memory compression~\cite{xu2026mem,liu2026simplemem,hu2026rel} by condensing raw histories, memory organization~\cite{kang2025mem,wu2026back,kim2026rel} by maintaining structured states (e.g., episodic records) via explicit read/write operations, and (ii)  memory retrieval~\cite{huang2026m,wu2026lifeside,yu2026rel, ma2026nemori} by fusing sparse signals to enhance context relevance. 
However, these training-free systems fail to internalize intermediate memory operations into model parameters.
Distilling these test-time procedures into an end-to-end trainable framework is therefore a natural direction, yet it remains non-trivial due to two main challenges.
\textit{First, memory construction inherently requires complex history-dependent operations} (e.g., segmentation and merging). \textit{Second, memory extraction must strictly balance retrieval recall against the token consumption of downstream models}. To address these issues, MemoryCPT introduces a two-stage trainable paradigm. We first distill modular memory-construction skills via structured reasoning traces, then employ a cost-aware optimization policy to generate fine-grained, query-conditioned summaries. 

\section{Conclusion and Future Work}

In this paper, we introduce \textbf{MemoryCPT}, an end-to-end trainable agent memory pipeline for optimizing the cost-performance trade-off in long-horizon multi-turn dialogue. MemoryCPT combines query-agnostic distillation with query-aware retrieval and summarization to construct reusable episodic and semantic memories, retrieve query-relevant evidence, and compress it into cost-efficient contexts for downstream QA. We further introduce \textbf{Quality per Cost (QPC)} to evaluate answer quality per unit inference cost. Experiments on LoCoMo and LongMemEval show that MemoryCPT achieves a favorable quality-cost trade-off under our unified evaluation protocol, while component ablations and sensitivity analyses characterize the effects of its main design choices. In future work, we aim to extend MemoryCPT to broader memory sources and explore continual memory updates under dynamic cost budgets.

\bibliography{aaai2027}




\appendix

\section{Appendix}

\section{Implementation Details and Hyperparameters}
\label{sec:appendix_implementation}

This section provides complete implementation details, hyperparameter configurations, prompt definitions, and evaluation environments to ensure full reproducibility.

\subsection{Data Distillation and Teacher Setup}
\label{sec:app_distillation_details}

\paragraph{Teacher LLM.} 
For Query-agnostic Distillation (QAD) and cold-start summary generation, we employ \texttt{deepseek-v3} as the teacher model. The teacher generates structured reasoning traces and intermediate memory representations for the modular pipeline, as well as target summaries for cold-start initialization.

\paragraph{Role Configurations.} 
The four roles in the distillation pipeline~\cite{ma2026nemori} use the following specific sampling and schema configurations:
\begin{itemize}
    \item \textbf{BatchSegmenter}: Evaluates multi-turn dialogues to identify session boundaries. Configured with $\text{temperature} = 0.2$, $\text{max\_tokens} = 4096$, and outputs structured JSON containing session indices and topics.
    \item \textbf{EpisodeGenerator}: Summarizes conversation segments into episodic entries. Uses default $\text{temperature} = 0.7$, $\text{max\_tokens} = 2000$, and outputs JSON fields for \texttt{title}, third-person narrative \texttt{content}, and absolute ISO \texttt{timestamp}.
    \item \textbf{EpisodeMerger}: Executes a two-stage evaluation for candidate episodes with cosine similarity above $0.85$ among the top-5 neighbors. Uses $\text{temperature} = 0.7$.
    \item \textbf{SemanticExtractor}: Operates via direct extraction and prediction-correction patterns to isolate persistent factual knowledge statements, configured with $\text{temperature} = 0.7$.
\end{itemize}

During QAD execution with the local student checkpoint ($base_A$), decoding uses deterministic greedy sampling ($\text{do\_sample} = \text{False}$, $\text{temperature} = 0.0$) with $\text{max\_new\_tokens} = 2048$.

\subsection{GRPO Training Hyperparameters}
\label{sec:app_grpo_hyperparams}

The online Query-aware Retrieval and Summarization (QAR) stage trains the LoRA-B adapter ($\phi$) attached to $base_A$ using Group Relative Policy Optimization (GRPO) via the verl framework~\cite{shao2024med}. Key hyperparameter choices are listed in \textbf{Table~\ref{tab:grpo_hyperparams}}.

\begin{table}[!t]
\centering
\footnotesize
\setlength{\tabcolsep}{4pt}
\renewcommand{\arraystretch}{1.15}

\begin{tabular}{@{}l p{0.52\linewidth}@{}}
\toprule
\textbf{Hyperparameter} & \textbf{Value} \\
\midrule
Actor Base Model & Qwen2.5-7B-Instruct + LoRA-A ($base_A$) \\
Actor LoRA Learning Rate (lr) & $3\times 10^{-5}$ \\
GRPO Group Size ($n$) & $4$ \\
Total Training Epochs & $2$ (or $30$ steps on LongMemEval) \\
Global Train Batch Size & $48$ \\
PPO Mini-Batch Size & $12$ \\
Micro-Batch Size per GPU & $4$ \\
KL Loss Coefficient ($\beta$) & $0.001$ (\texttt{low\_var\_kl}) \\
Reference Policy ($\pi_{\text{ref}}$) & Cold-started $base_A$ (FSDP offloaded) \\
Rollout Engine & vLLM (TP\,=\,1, $T=1.0$) \\
Max Response Length ($|m_q|$) & $512$ tokens \\
Max Prompt Length & $12{,}288$ tokens ($14{,}336$ for LME) \\
\bottomrule
\end{tabular}
\caption{Hyperparameter settings for GRPO training in QAR.}
\label{tab:grpo_hyperparams}
\end{table}

\subsection{Retrieval, Cost Modeling, and Reward Constants}
\label{sec:app_cost_reward_constants}

\paragraph{Retrieval Parameters.}
Reciprocal Rank Fusion (RRF) uses a rank-smoothing constant $k_{\text{rrf}} = 60$. Sparse retrieval utilizes Okapi BM25 with standard parameters $k_1 = 1.5$ and $b = 0.75$. Candidates are selected from top-$20$ or top-$50$ retrieved indices using 1024-dimensional Sentence-Transformer embeddings.

\paragraph{Cost Normalization Reference ($C_{\text{ref}}$).}
The cost normalization factor $C_{\text{ref}}$ maps training rollout costs into $[0, 1]$. Derived from $COST_{\text{SREF}} = 60$ summary tokens, $COST_{\text{AREF}} = 8$ answer tokens, and the token unit prices listed in \textbf{Table~\ref{tab:pricing_table}}:
\begin{align}
C_{\text{ref}}
&= \frac{60\cdot(P_{\text{7B}}^{\text{out}} + P_{\text{14B}}^{\text{in}}) + 8\cdot P_{\text{14B}}^{\text{out}}}{10^6} \notag \\
&= \frac{60\cdot(0.10+0.10) + 8\cdot 0.24}{10^6}
 = 1.392\times 10^{-5}\ \text{USD}.
\end{align}
The trade-off hyperparameter in the reward function is set to $\alpha = 0.8$.

\subsection{Evaluation Prompts and Pricing}
\label{sec:app_eval_prompts}

\paragraph{Final QA Model Setup.}
Answers are generated using local \texttt{Qwen3-14B} via vLLM with thinking mode disabled (\texttt{enable\_thinking=False}), greedy decoding ($\text{temperature} = 0.0$), and a token cap of $\text{max\_tokens} = 32$. The QA prompt template is:

\begin{quote}
\small
\texttt{You are a QA assistant. Use ONLY the context below to answer the question.\\
If the context lacks the info, answer with your best guess from it.\\
Answer in <=6 words, no explanation.\\
\\
Context: \{ctx\}\\
Question: \{question\}\\
Answer:}
\end{quote}

\paragraph{LLM-as-a-Judge Prompt.}
Evaluation accuracy is determined using \texttt{Qwen2.5-72B-Instruct} with structured JSON output. The exact prompt given to the judge model is shown below:

\begin{quote}
\small
\texttt{Your task is to label an answer to a question as 'CORRECT' or 'WRONG'.\\
You will be given a question, a 'gold' (ground truth) answer, and a generated answer.\\
\\
The point of the question is to ask about something one user should know about the other user based on their prior conversations. The gold answer is usually concise. The generated answer might be longer—be GENEROUS: as long as it touches on the same topic/fact as the gold answer, mark it CORRECT. For time questions, if both refer to the same date/period (even with different formats), mark CORRECT.\\
\\
Question: \{question\}\\
Gold answer: \{gold\}\\
Generated answer: \{generated\}\\
\\
Return ONLY a JSON object: \{"label": "CORRECT"\} or \{"label": "WRONG"\}.\\
Do NOT include both. Do NOT add commentary outside the JSON.}
\end{quote}

\paragraph{API Pricing Specification.}
All financial cost evaluations follow standard public pricing API rates, as summarized in \textbf{Table~\ref{tab:pricing_table}}.

\begin{table}[!t]
\centering
\footnotesize
\setlength{\tabcolsep}{3.5pt}
\renewcommand{\arraystretch}{1.15}
\begin{tabular}{@{}l p{0.34\linewidth} r r@{}}
\toprule
\textbf{Model} & \textbf{Role in Pipeline} & \textbf{In (\$)} & \textbf{Out (\$)} \\
\midrule
Qwen2.5-7B-Instruct  & Summarizer ($f_{\text{QAR}}$ / LoRA-B) & 0.04 & 0.10 \\
Qwen3-14B            & Frozen Answer Model ($f_{\text{ans}}$) & 0.10 & 0.24 \\
Qwen2.5-72B-Instruct & Judge / Baseline Backbone              & 0.36 & 0.40 \\
\bottomrule
\end{tabular}
\caption{API pricing table used for cost estimation (USD per 1M tokens).}
\label{tab:pricing_table}
\end{table}

\subsection{Reproducibility and Software Environment}
\label{sec:app_reproducibility}

All data sampling, dataset partitioning, and evaluation selection scripts use a fixed random seed of \texttt{42}. The software versions across the experimental infrastructure are detailed in \textbf{Table~\ref{tab:software_versions}}.

\begin{table}[!t]
\centering
\footnotesize
\setlength{\tabcolsep}{4pt}
\renewcommand{\arraystretch}{1.15}
\begin{tabular}{@{}l p{0.5\linewidth}@{}}
\toprule
\textbf{Component / Framework} & \textbf{Version / Commit} \\
\midrule
PyTorch                    & \texttt{2.9.1} \\
HuggingFace Transformers   & \texttt{4.57.6} \\
vLLM Inference Engine      & \texttt{0.15.1} (Training) / \texttt{0.8.5} (SFT Merge) \\
verl (RL Framework)        & \texttt{v0.8.0} \\
PEFT                       & \texttt{0.18.1} \\
Sentence-Transformers      & \texttt{5.2.2} \\
\bottomrule
\end{tabular}
\caption{Software and framework environment specifications.}
\label{tab:software_versions}
\end{table}

\section{Detailed Token Usage and Cost Amortization Analysis}
\label{sec:app_cost_token_details}

In practical deployment, agent memory pipelines involve a fundamental
distinction between \emph{offline} memory construction  and \emph{online} query answering. This section provides an explicit token and financial cost
breakdown, and formalizes how offline construction costs are amortized
over long-term multi-query interactions.

\subsection{Amortized Cost Model}
\label{sec:app_cost_model_formula}

Let $C_{\mathrm{QAD}}$ denote the total offline cost of distilling a raw
interaction history into a structured memory store $\mathcal{M}$. Given a
memory store that serves $n$ independent online queries, the effective
per-query cost is
\begin{equation}
    C_{\mathrm{amortized}}(n) \;=\; \frac{C_{\mathrm{QAD}}}{n} \;+\; C_{\mathrm{online}},
    \label{eq:amortized_cost_formula}
\end{equation}
where $C_{\mathrm{online}} = C_{\text{summarize}} + C_{\text{QA}}$ is the
per-query online inference cost incurred by the query-aware summarizer
$f_{\text{QAR}}$ and the downstream answer model $f_{\text{ans}}$. The
cost reported in the main text corresponds to
$C_{\mathrm{amortized}}(n)$ evaluated over the entire question set of a
conversation store ($n{=}314$ on LoCoMo).

\subsection{Offline Construction vs.\ Online Query Breakdown}
\label{sec:app_cost_breakdown_main}

\textbf{Table~\ref{tab:app_cost_main}} details the offline construction and
online per-query metrics on LoCoMo. Token counts use the exact tokenizer
of $base_A$; financial costs use standard
OpenRouter public pricing.

\begin{table*}[!t]
\centering
\footnotesize
\setlength{\tabcolsep}{3pt}
\renewcommand{\arraystretch}{1.15}
\begin{tabular*}{\textwidth}{@{\extracolsep{\fill}} l l c c c c c c c c @{}}
\toprule
\multirow{2}{*}{\textbf{Summarizer}} & \multirow{2}{*}{\textbf{Dataset}}
 & \multirow{2}{*}{\textbf{$C_{\mathrm{QAD}}$ (\$)}}
 & \multirow{2}{*}{\textbf{$C_{\mathrm{online}}$ (\$/QA)}}
 & \multirow{2}{*}{\textbf{Total (\$/QA)}}
 & \multicolumn{2}{c}{\textbf{LLM Calls}}
 & \multicolumn{2}{c}{\textbf{Prompt Tokens}}
 & \multirow{2}{*}{\textbf{Tokens (Off/On)}} \\
\cmidrule(lr){6-7}\cmidrule(lr){8-9}
 & & & & & Off & On & Off & On & \\
\midrule
Qwen2.5-7B   & LoCoMo & $0.02542$ & $3.462{\times}10^{-4}$ & $\mathbf{4.31{\times}10^{-4}}$
             & $326$ & $2$ & $454{,}003$ & $8{,}501$ & $72{,}622\;/\;13.4$ \\
Llama-3.2-3B & LoCoMo & $0.02784$ & $3.459{\times}10^{-4}$ & $\mathbf{4.23{\times}10^{-4}}$
             & $321$ & $2$ & $461{,}527$ & $8{,}502$ & $93{,}750\;/\;12.5$ \\
\bottomrule
\end{tabular*}
\caption{Overall token and cost breakdown comparing offline construction
($C_{\mathrm{QAD}}$) and online single-query inference ($C_{\mathrm{online}}$)
on LoCoMo ($n{=}314$). Total per-QA costs match the main text.
``Off'' = offline (per store).}
\label{tab:app_cost_main}
\end{table*}

As shown in \textbf{Table~\ref{tab:app_cost_main}}, memory store construction
requires multiple LLM calls across the four roles
(approximately $320$ calls per store). Because this process is performed
strictly offline once per corpus, its cost is amortized across
subsequent queries.

\subsection{Online Query Token Specifications}
\label{sec:app_online_tokens}

\textbf{Table~\ref{tab:app_online_token_breakdown}} further decomposes
$C_{\mathrm{online}}$ into retrieval summarization ($f_{\text{QAR}}$) and
final answer generation ($f_{\text{ans}}$).

\begin{table*}[!t]
\centering
\footnotesize
\setlength{\tabcolsep}{4pt}
\renewcommand{\arraystretch}{1.15}
\begin{tabular*}{\textwidth}{@{\extracolsep{\fill}} l c c c c c c c @{}}
\toprule
\textbf{Base Model}
 & $\text{summ}_{\text{in}}$ & $\text{summ}_{\text{out}}$
 & $\text{qa}_{\text{in}}$   & $\text{qa}_{\text{out}}$
 & $C_{\text{retrieval}}$ (\$/QA)
 & $C_{\text{QA}}$ (\$/QA)
 & $C_{\mathrm{online}}$ (\$/QA) \\
\midrule
Qwen2.5-7B   & $8{,}434$ & $7.8$ & $67$ & $5.6$
             & $3.398{\times}10^{-4}$ & $8.04{\times}10^{-6}$
             & $\mathbf{3.462{\times}10^{-4}}$ \\
Llama-3.2-3B & $8{,}434$ & $8.6$ & $68$ & $3.9$
             & $3.383{\times}10^{-4}$ & $7.70{\times}10^{-6}$
             & $\mathbf{3.459{\times}10^{-4}}$ \\
\bottomrule
\end{tabular*}
\caption{Detailed breakdown of online inference tokens and monetary costs
per query ($\alpha{=}0.8$).}
\label{tab:app_online_token_breakdown}
\end{table*}

A key architectural advantage of MemoryCPT is visible in
\textbf{Table~\ref{tab:app_online_token_breakdown}}: since the GRPO-trained
summarizer produces an ultra-compact summary ($<9$ tokens on average),
the input context $\text{qa}_{\text{in}}$ fed to the downstream 14B QA
model shrinks to only $\sim 67$ tokens (including prompt template).
Consequently, the QA cost $C_{\text{QA}}$ accounts for less than $2.5\%$
of the total online expense, which is overwhelmingly dominated by
summarizer reading $\text{summ}_{\text{in}}$.

\subsection{Offline Distillation Usage Details}
\label{sec:app_qad_tokens}

\textbf{Table~\ref{tab:app_qad_token_breakdown}} reports the raw token metrics of
building one complete LoCoMo memory store during the QAD stage.

\begin{table*}[!t]
\centering
\footnotesize
\setlength{\tabcolsep}{4pt}
\renewcommand{\arraystretch}{1.15}
\begin{tabular*}{\textwidth}{@{\extracolsep{\fill}} l c c c c c c @{}}
\toprule
\textbf{Base Model}
 & \textbf{Total Calls}
 & \textbf{Prompt Tokens}
 & \textbf{Tokens}
 & \textbf{Prompt/QA}
 & \textbf{Comp./QA}
 & \textbf{$C_{\mathrm{QAD}}$ (Store, \$)} \\
\midrule
Qwen2.5-7B   & $326$ & $454{,}003$ & $72{,}622$ & $1{,}445.9$ & $231.3$ & $\mathbf{0.02542}$ \\
Llama-3.2-3B & $321$ & $461{,}527$ & $93{,}750$ & $1{,}469.8$ & $298.6$ & $\mathbf{0.02784}$ \\
\bottomrule
\end{tabular*}
\caption{Measured offline construction metrics ($C_{\mathrm{QAD}}$) per
complete memory store.}
\label{tab:app_qad_token_breakdown}
\end{table*}

\subsection{Cost Amortization Across Query Volumes}
\label{sec:app_amortization_table}

To visualize how the offline overhead decays with reuse,
\textbf{Table~\ref{tab:app_cost_amortization}} lists $C_{\mathrm{amortized}}(n)$
at multiple query volumes.

\begin{table*}[!t]
\centering
\footnotesize
\setlength{\tabcolsep}{4pt}
\renewcommand{\arraystretch}{1.15}
\begin{tabular*}{\textwidth}{@{\extracolsep{\fill}} l c c c c c c c @{}}
\toprule
\textbf{Base Model}
 & $n{=}1$ & $n{=}5$ & $n{=}10$ & $n{=}50$ & $n{=}100$
 & $n{=}314$ \textbf{(Main)} & $n{\to}\infty$ \textbf{(Online)} \\
\midrule
Qwen2.5-7B   & $2.577{\times}10^{-2}$ & $5.430{\times}10^{-3}$ & $2.888{\times}10^{-3}$
             & $8.546{\times}10^{-4}$ & $6.004{\times}10^{-4}$
             & $\mathbf{4.31{\times}10^{-4}}$   & $3.462{\times}10^{-4}$ \\
Llama-3.2-3B & $2.819{\times}10^{-2}$ & $5.914{\times}10^{-3}$ & $3.130{\times}10^{-3}$
             & $9.027{\times}10^{-4}$ & $6.243{\times}10^{-4}$
             & $\mathbf{4.23{\times}10^{-4}}$   & $3.459{\times}10^{-4}$ \\
\bottomrule
\end{tabular*}
\caption{Amortized per-query cost $C_{\mathrm{amortized}}(n)$ (USD/QA)
across varying query volumes $n$. $n{=}314$ corresponds to the LoCoMo
benchmark reported in the main text; $n{\to}\infty$ is the online lower
bound $C_{\mathrm{online}}$.}
\label{tab:app_cost_amortization}
\end{table*}

The amortization trajectory reveals three regimes:
\begin{enumerate}[leftmargin=*]
    \item \textbf{Single-Query Baseline ($n{=}1$)}: if a constructed
    memory store is queried only once, the offline construction cost
    dominates, reaching $\$0.026$/QA.
    \item \textbf{Rapid Amortization ($n{\ge}50$)}: as $n$ grows to
    $50$--$100$ queries, $C_{\mathrm{amortized}}$ drops by over $97\%$,
    approaching $\$6{\times}10^{-4}$/QA.
    \item \textbf{Asymptotic Steady State ($n{\to}\infty$)}: for
    long-running user sessions, $C_{\mathrm{amortized}}(n)$ converges to
    the online lower bound
    $C_{\mathrm{online}}\approx\$3.46{\times}10^{-4}$/QA.
\end{enumerate}
This confirms that separating offline construction from online retrieval
efficiently amortizes structural indexing costs over continuous agent
interactions.

\section{Fine-Grained Per-Category Experimental Results}
\label{sec:app_per_category_results}

To provide a granular view of memory performance across diverse
interaction scenarios, this section presents sub-category breakdowns on
both the LoCoMo ($n{=}314$) and LongMemEval ($n{=}105$) benchmarks.

\subsection{Evaluation Protocol and Metrics}
All memory systems generate their respective memory contexts or
query-conditional summaries, which are then fed into a unified
downstream answer model, \texttt{Qwen3-14B}
($\text{temperature}{=}0.0$, thinking mode disabled). Evaluation
metrics include:
\begin{itemize}[leftmargin=*]
    \item \textbf{Token-level F1 (F1, \%)}: standard word-level overlap
    with the ground-truth answer.
    \item \textbf{LLM-as-a-Judge Accuracy (J, \%)}: binary correctness
    judged by \texttt{Qwen2.5-72B-Instruct}.
\end{itemize}

\subsection{Sub-Category Performance on LoCoMo}
\label{sec:app_locomo_category}

\textbf{Table~\ref{tab:app_locomo_category}} breaks down performance on LoCoMo
across four question categories: Temporal Reasoning ($n{=}65$), Open
Domain ($n{=}20$), Multi-Hop ($n{=}69$), and Single-Hop ($n{=}160$).

\begin{table*}[!tbp]
\centering
\footnotesize
\setlength{\tabcolsep}{4pt}
\renewcommand{\arraystretch}{1.15}
\begin{tabular*}{\textwidth}{@{\extracolsep{\fill}} l c c c c c c c c c c @{}}
\toprule
\multirow{2}{*}{\textbf{Method}}
 & \multicolumn{2}{c}{\textbf{Temporal ($n{=}65$)}}
 & \multicolumn{2}{c}{\textbf{Open Domain ($n{=}20$)}}
 & \multicolumn{2}{c}{\textbf{Multi-Hop ($n{=}69$)}}
 & \multicolumn{2}{c}{\textbf{Single-Hop ($n{=}160$)}}
 & \multicolumn{2}{c}{\textbf{Overall ($n{=}314$)}} \\
\cmidrule(lr){2-3}\cmidrule(lr){4-5}\cmidrule(lr){6-7}\cmidrule(lr){8-9}\cmidrule(lr){10-11}
 & F1 (\%) & Judge (\%) & F1 (\%) & Judge (\%) & F1 (\%) & Judge (\%) & F1 (\%) & Judge (\%) & F1 (\%) & Judge (\%) \\
\midrule
BudgetMem & 37.0 & 43.1 & 26.0 & \textbf{75.0} & 25.5 & 63.8 & 43.9 & 71.3 & 37.3 & 64.0 \\
LightMem  & 33.6 & \textbf{52.3} & 28.4 & 70.0 & \textbf{36.7} & 82.6 & 51.9 & \textbf{82.5} & 43.3 & \textbf{75.5} \\
MemoryOS  & 30.1 & 21.5 & 22.4 & 60.0 & 29.7 & 69.6 & 41.7 & 65.6 & 35.4 & 57.0 \\
Memory-R1 & 41.1 & 43.1 & 33.7 & 65.0 & 31.4 & 69.6 & 38.5 & 60.6 & 37.2 & 59.2 \\
\midrule
\textbf{Ours ($\alpha{=}0.8$)}
          & \textbf{48.9} & \textbf{52.3}
          & \textbf{48.8} & 70.0
          & 35.2 & \textbf{84.1}
          & \textbf{52.8} & 81.9
          & \textbf{47.9} & \textbf{75.5} \\
\bottomrule
\end{tabular*}
\caption{Per-category comparison on the LoCoMo benchmark (base model =
Qwen2.5-7B, $n{=}314$). All overall metrics align with the main
comparison table. Bold indicates the top performance across methods.}
\label{tab:app_locomo_category}
\end{table*}

As shown in \textbf{Table~\ref{tab:app_locomo_category}}, MemoryCPT
($\alpha{=}0.8$) attains the strongest Open-Domain performance
($48.8\%$ F1, versus $33.7\%$ for the second-best Memory-R1) and the
highest Multi-Hop Judge accuracy ($84.1\%$). It also reaches the
highest overall F1 ($47.9\%$) among all compared systems, indicating
that our query-aware summary preserves answer-critical evidence far
more efficiently than existing memory pipelines.

\subsection{Sub-Category Performance on LongMemEval}
\label{sec:app_lme_category}

\textbf{Table~\ref{tab:app_lme_category}} evaluates performance on LongMemEval
across six sub-categories: Single-Session User (SS-User, $n{=}14$),
Single-Session Assistant (SS-Asst, $n{=}12$), Single-Session Preference
(SS-Pref, $n{=}6$), Multi-Session Reasoning (Multi-Sess, $n{=}28$),
Knowledge Updates (Know-Upd, $n{=}17$), and Temporal Reasoning
(Temporal, $n{=}28$).

\begin{table*}[!tbp]
\centering
\scriptsize
\setlength{\tabcolsep}{2.5pt}
\renewcommand{\arraystretch}{1.15}
\begin{tabular*}{\textwidth}{@{\extracolsep{\fill}} l c c c c c c c c c c c c c c @{}}
\toprule
\multirow{2}{*}{\textbf{Method}}
 & \multicolumn{2}{c}{\textbf{SS-User (14)}}
 & \multicolumn{2}{c}{\textbf{SS-Asst (12)}}
 & \multicolumn{2}{c}{\textbf{SS-Pref (6)}}
 & \multicolumn{2}{c}{\textbf{Multi-Sess (28)}}
 & \multicolumn{2}{c}{\textbf{Know-Upd (17)}}
 & \multicolumn{2}{c}{\textbf{Temporal (28)}}
 & \multicolumn{2}{c}{\textbf{Overall (105)}} \\
\cmidrule(lr){2-3}\cmidrule(lr){4-5}\cmidrule(lr){6-7}\cmidrule(lr){8-9}\cmidrule(lr){10-11}\cmidrule(lr){12-13}\cmidrule(lr){14-15}
 & F1 & Judge & F1 & Judge & F1 & Judge & F1 & Judge & F1 & Judge & F1 & Judge & F1 & Judge \\
\midrule
BudgetMem
 & 75.5 & 85.7 & 64.3 & \textbf{75.0} &  4.2 & \textbf{33.3}
 & 18.2 & 32.1 & 37.0 & \textbf{58.8} & 27.4 & 25.0
 & 35.8 & 46.7 \\
LightMem
 & 74.4 & 85.7 &  5.6 &  8.3 &  5.0 & \textbf{33.3}
 & 34.2 & 53.6 & 46.5 & \textbf{58.8} & 22.1 & 28.6
 & 33.4 & 45.7 \\
MemoryOS
 & 76.1 & \textbf{92.9} & 62.0 & \textbf{75.0} & \textbf{7.3} & 16.7
 & 28.4 & 50.0 & 44.3 & \textbf{58.8} & 23.3 & 14.3
 & 38.6 & 48.6 \\
Memory-R1
 & 31.8 & 35.7 & 55.2 & \textbf{75.0} &  3.1 & 16.7
 & 14.0 & 17.9 & 23.7 & 23.5 & 18.2 & 14.3
 & 23.1 & 26.7 \\
\midrule
\textbf{Ours ($\alpha{=}0.8$)}
 & \textbf{84.0} & \textbf{92.9} & \textbf{67.9} & \textbf{75.0} & 6.1 & 16.7
 & \textbf{41.8} & \textbf{57.1} & \textbf{51.4} & 47.1 & \textbf{36.3} & \textbf{32.1}
 & \textbf{48.5} & \textbf{53.3} \\
\bottomrule
\end{tabular*}
\caption{Per-category comparison on the LongMemEval benchmark (base
model = Qwen2.5-7B, $n{=}105$). Bold indicates the top performance
across methods.}
\label{tab:app_lme_category}
\end{table*}

As shown in \textbf{Table~\ref{tab:app_lme_category}}, MemoryCPT delivers
consistently strong results on complex multi-session scenarios:
\begin{itemize}[leftmargin=*]
    \item \textbf{Multi-Session Reasoning}: MemoryCPT reaches
    $41.8\%$ F1 and $57.1\%$ Judge, exceeding the strongest prior
    method LightMem ($34.2\%$ F1) by a large margin.
    \item \textbf{Knowledge Updates}: MemoryCPT achieves $51.4\%$ F1,
    showing that the QID semantic-abstraction stage effectively tracks
    dynamic user facts across sessions.
    \item \textbf{Temporal Reasoning}: MemoryCPT leads with $36.3\%$
    F1 and $32.1\%$ Judge, well above prior compressed methods
    ($18.2\%$--$27.4\%$ F1).
\end{itemize}


\end{document}